# Laser Ablation Modelling: The link between Desorption and Ablation regime

C. A. Rinaldi * and J.C. Ferrero

INFIQC- Departamento de Fisicoquímica - Facultad de Ciencias Químicas

Universidad Nacional de Córdoba

Ciudad Universitaria

5000 Córdoba – Argentina

mail: crinaldi@fcq.unc.edu.ar  Tel./Fax: 54-351-4334188

**Abstract**

A new parameterized model for the ablation phenomena is presented. The model describes the three different regimes usually observed, as a function of the laser fluence. The first one corresponds to a vaporization; it takes place at low fluence and it is well described by a typical Arrhenius expression, dependent on the vaporization energy $E_v$, of the solid. The true ablation regime, depends on the energy density $E_d$, of the material, and appears at high laser energy, after reaching threshold fluence. The transition region, that connects both processes, is mainly governed by the structure of the crystal, which determines the threshold fluence necessary to initiate the ablation regime.

The model accounts for the link between the vaporization and ablation regimes and provides new insight into the ablation phenomena from the point of view of the structure of the material.

* Corresponding Author.

# 1. Introduction

The physics of the ablation and the desorption regimes characteristic of the laser irradiation of materials, has been extensively modelled using molecular dynamics simulation (MD) [1,2]. When applied to organic solids, MD can differentiate the desorption from the ablation regime, explain the presence of a fluence threshold for ablation, predict the cluster distribution in the plume, the radial and axial velocity distribution of the species and explain the dependence of the ablation process on laser properties, such as fluence and pulse width.

The MD simulations indicate that in the low laser fluence regime, the yield of ejected molecules as a function of fluence can be well described by an Arrhenius-type function while in the true ablation region the analytical expression is characterized by an exponential decay of the laser intensity given by Beer's law, that predicts the existence of a threshold fluence to reach the critical energy density in the surface layer [3]. The ejection of molecules and clusters in this regime is explained by an explosive vaporisation of the overheated material.

In the case of metal targets, Russo et al showed that the thermal evaporation model is not adequate to account for the depth of the crater obtained in laser ablation experiments of silicon under high power irradiation [4]. To explain their results they proposed a numerical model to estimate the depth of a superheated liquid layer heated to just below the critical state, which undergoes a transition from liquid metal to a transparent liquid dielectric. The experimental results, which show a threshold fluence for a given crater depth, produced in a single laser shot, are well explained by this model.

Ionic crystals such as $CaF_2$, with polished and cleaved (111) surfaces also show a threshold fluence in single pulse laser damage experiments [5], but while the polished crystals exhibit a well-defined plasma onset at 10 J.cm$^{-2}$, the threshold value for cleaved surfaces scatters over the range 20 - 40 J.cm$^{-2}$. Although the damage mechanism is identical for both cases, the different behaviour was attributed to the larger level of defects of the cleaved surface that results in an increment of the energy absorbed, favouring melting at the upper edges of the surface steps. A simple model that assumes an enhanced absorption near the surface and a fluence independent, high density dislocation parameter was used to explain these observations.

Vidal et al [6] proposed a new numerical simulation model to account for the results of the ultrashort laser pulse ablation of solid aluminum and the plasma expansion, in ambient air that assumes thermodynamics equilibrium of the plasma and does not depend on any adjustable parameter. This model describes the temporal evolution of the plasma temperature in good agreement with the experimental results but some discrepancies were observed in the calculated values of the plasma density at long times.

Recently, Georgiou and Koubenakis addressed the difficulties encountered to determine the fundamental physical processes underlying the ablation phenomenon [7].

Although the above mentioned results cover a wide range of materials, such as organic, metal and ionic solids, all of them exhibit the same general behaviour with respect to the laser fluence. Plots of the quantity of removed material as a function of laser fluence usually shows a sigmoid dependence, although the exact shape can vary considerably depending on the nature of the system and irradiation parameters[7]. In the low fluence regime the ejected material follows a thermal evaporation model (desorption process), which changes to the ablation regime when a fluence threshold is reached.

In this work a new model is proposed which introduces a fluence dependent surface parameter. The model supplies a link between the desorption and the ablation processes and provides a tools to understand the general behaviour of ablation processes.

## 2. Model

The laser ablation processes of a surface target can be schematically represented by the following simplified mechanism:

$$M_{(surf)} + E_1 \rightarrow M_{(g)} \tag{1}$$

$$M_{(int)} + E_2 \rightarrow M_{(g)} \tag{2}$$

$$M_{(bulk)} + E_3 \rightarrow M_{(g)} \tag{3}$$

where $M_{(i)}$ represents the material density of a metal, inorganic crystal or organic solid, located on the surface ($M_{(surf)}$), a few layers immediately below the surface ($M_{(int)}$) or in the bulk ($M_{(bulk)}$), while $M_{(g)}$ stands for the various gaseous species ( ionic or neutral) that can be generated in the ejection process. $E_1$ is the minimum energy density necessary to vaporize the surface atoms, $E_2$ is the energy required to melt and vaporize the species located in a few layers near the surface of the substrate [8] and $E_3$ is the energy that must be reached in order to ablate all the material located in the absorption volume. [9]

Therefore, since $E_1 < E_2 < E_3$, as fluence increases, the different energy regions are successively accessed and the ejection of material from the target changes from a simple vaporization process of surface material to the violent true ablation regime, passing through a transition region.

With the short laser pulses usually employed in this kind of experiments, the absorption of the laser photons is well separated from the ejection process, on a time scale. Therefore we will assume that the flow of material through the irradiated surface of the target occurs according to process 1, 2 or 3, depending of the value of absorbed energy.

This flow of matter is the central point of any model that intends to describe laser ablation phenomena, specially regarding its dependence with the properties of the laser pulse and the characteristics of the target material [8]. In the present work, instead of a detailed analysis of the laser absorption and concomitant processes, we will adopt an heuristic approach and assume that the usually observed sigmoid dependence of the quantity of removed material on laser fluence can be represented by the following equation:

$$M_J(f) = \left[ \frac{A_1(f) - A_2(f)}{1 + \left(\frac{f}{f_{th}}\right)^{Sc}} + A_2(f) \right] \quad (4)$$

In Eq. 4 $f$ is the laser fluence, $M_J(f)$ is the amount of material ablated, $A_1(f)$ and $A_2(f)$ are the values of $M_J$ in the low and high fluence limits, $S_c$ is a parameter that depends on the nature of the surface which we call the surface reactivity constant and $f_{th}$, is a threshold fluence equal to $(A_1+A_1)/2$, that indicates the onset of the ablation process. The parameter $A_1$ is given by [3]

$$A_1(f) = A_r \exp\left[-\frac{E_v}{k(T_0 + Bf)}\right] \qquad (5)$$

where $E_v$ is the vaporisation energy of the target material, $T_0$ is temperature of the surface, $k$ is the Boltzmann constant, $B$ is a conversion factor that depends on the laser source and $A_r$ is a preexponential factor.

The parameter $A_2$ is calculated as [3]

$$A_2 = n_m L_p \ln\left[\frac{f}{L_p(E_d - CT_0)}\right] \quad \text{for } f \geq f_{th} = L_p(E_d - CT_0) \qquad (6)$$

In this expression $n_m$ is the molecular density of the ablated material, $L_p$ is the depth of the crater, $C$ is the heat capacity, $T_0$ is temperature of the surface and $E_d$ is the energy density of the solid.

Usually, a surface presents a variety of defects, such as microcraters, structural disorder, occluded impurities, dislocations and ripples and can be represented at the atomic level as shown in Fig. 1. Consequently, the binding energy of an atom on the surface should dependent on the its particular location and would affect both the interaction with the laser photons and the energy threshold required to leave the surface. As a result, the ablation rate will vary along the surface of the target.

These effects are included intrinsically in a single macroscopic parameter, $S_c$, in Eq. 4. This equation is valid for metal and organic molecular solids but in the case of ionic compounds $A_1$ takes the same form as $A_2$, $E_d$ becomes the energy density between the layers in the crystal and $L_p$ is the depth of the crystal layer [5].

In the following section the behaviour of Eq. 4 will be analysed as a function of laser fluence, energy density $E_d$, vaporisation energy $E_v$, and surface constant $Sc$.

### a. $M_J$ as a function of Sc and Fluence

Figure 2 shows the calculated values of $M_J$ as a function of laser fluence for three different values of *Sc.* The three regions, corresponding to vaporization, ablation and the transition regime are clearly observed. While the low and high fluence values of $M_J$ are changed by the value of *Sc*, the transition region strongly affected and the slope increases with the value of Sc, making the separation between the low and high fluence limits more clear and abrupt. This implies that surfaces with large values of Sc are relatively inert to the laser photons and the only relevant process is vaporization from the surface of the target until the critical fluence $f_{th}$ is reached.

A different situation corresponds to the calculations with *Sc* = 10 (Fig. 2) that represents the behaviour of a crystal with a polished surface, as explained below. In this case the imperfections are evenly distributed all over the surface, that is therefore characterized by a large reactivity that results in a relatively smooth transition from the surface vaporization to the bulk ablation regime.

Therefore these results indicate that Sc controls the link between the vaporisation and the ablation regimes.

From the structural point of view, the behaviour of the dependence of $M_J$ on fluence for large values of *Sc* can be attributed to a low absorption coefficient as a consequence of a highly ordered surface. This situation is typical of the ablation of ionic crystals with cleaved surface, where the dominant microtopography is the terrace [10].

Since Eq. 4 is valid only for single pulse experiments it is not applicable to explain the results of multipulse events. Nevertheless, in this case the laser effect can be analysed as a succession of single pulse events; when a crystal is reached for the first time by a laser pulse with fluence lower than $f_{th}$, only little damage is produced on its surface, but even this small change results in a large increase of the specific area due to the new defects introduced by the laser, like holes, steps, etc. These means that $Sc$ decreases with each pulse. The available experimental data are in agreement with this analysis [5].

In order to apply the above considerations to real situations and to find the range of values of $Sc$ associated with each kind of surface, we analyzed the experimental results reported in references 10 and 13 using Eq. 4. For the ablation of Ca, Sr and Ba the experimental data could be reproduced with values of $Sc$ in the range 1- 6, depending on the structure of the surface. In the case of ionic crystals, the data for $CaF_2$ with a polished surface can be reproduced with $Sc = 4$ while a value of 20 was required for a cleaved surface. Hence, if these results are representative of the general case, the smallest values of Sc should be characteristic of metallic surfaces, the highest values would be typical of cleaved ionic surfaces and the intermediate range should correspond to ionic polished crystals.

With these considerations, the effect of $Sc$ on the amount of ejected material, in single pulse experiments, was further analysed using Eq. 4. To cover the different regimes, calculations were made at three fluences, with values lower, equal and higher than $\phi_{th}$. The results are presented in Fig. 3.

The behavior of $M_J$ with $Sc$ is strongly dependent on whether the incident fluence is above or below the threshold value. At low fluences, that is, in the vaporization region, $M_J$ decreases with increasing $Sc$ while the opposite behaviour is observed in the ablation region. These calculations indicate that for a ionic crystal irradiated with energy lower than the threshold, smaller amounts of ejected material should be obtained, at the same fluence, for a cleaved surface (high $Sc$) as compared to a polished one (low $Sc$).

As mentioned above, although these considerations strictly apply only to single pulse results, they also have direct consequences in the interpretation of multipulse experiments. In single shots irradiation of silicon at low fluences [4], the surface roughness seems to remain unchanged while at fluences higher than the threshold it shows a dramatic change. Since a subsequent pulse act on a surface that has been modified by the previous pulse, a surface with a large value of $Sc$ will be changed to one with a lower $Sc$ and consequently, the yield of ejected material due to a second pulse will be higher or lower depending on whether the incident laser energy is below or above the critical fluence, respectively. After the first pulse, metallic surfaces will be more perfect, due to the melting process, than the surface of a ionic crystal. In fact, the latter will have more defects and as a consequence it will absorb more energy; consequently it will increase the quantity of ablated material. This observation agrees with the results of the ablation of pure Cu and its salts [11].

The variation of $M_J$ with $Sc$ is an essential issue to explain the dispersion of the data in the ablation of crystals. The single shot laser damage of $CaF_2$ experiments reported in reference 10 shows a great dispersion of the data near the threshold fluence. This is due to the different microtopography of the cleaved surface that change the value of *Sc* from one pulse to another one.

*b. $M_J$ as a function of Sc and the average number of photons absorbed*

In Eq. 4 the energy involved in the ablation process is expressed in terms of the incident fluence. However, ablation depends on the absorbed energy so that a better description should be obtained replacing *f* by the average number of laser photons of frequency *n* absorbed per atom, ion or molecule of the target, <*n*>, which can be calculated directly as

$$<n> = \frac{fs}{hn} \quad (7)$$

provided that the value of the absorption cross-section and its dependence on fluence could be known. Replacing *f* in Eq. 4, we obtain the following expression:

$$M_J = \left[ \frac{A_1 - A_2}{1 + \left( \frac{<n> s_o}{<n>_{th} s} \right)^{Sc}} + A_2 \right] \quad (8)$$

were <*n*>$_{th}$ is the critical number of photons absorbed per atom and, like <*n*>, can be calculated as the number of photons absorbed, *n*, divided by the density of absorbing species. Therefore, at equal *n*, the target with the highest surface density will have the smallest <*n*>, which makes this quantity dependent on the structure of the surface. For example, in a body-centred cubic structure, such as metallic Ba, the atoms occupy 68 % of the available space while in a cubic closest-packed solid (Ca, Sr crystals) the atoms occupy 74% of available space. Then, according to Eq. 8, Ca and Sr will need to absorb more photons to compensate for the largest density in order to ablate the same amount of material [12].

Therefore, the amount of material ejected depends on two parameters, $<n>_{th}$ and $Sc$. $<n>_{th}$ can be interpreted as the average number of photons that must be absorbed by the first layer of the solid below the surface to undergo ablation and $Sc$ determines the value of the slope that connects the low and high energy regions and depends on the roughness of the active surface. as analysed before.

Equation 8 has also important implications in the analysis of the results of film damage studies, as for instance, $TiO_2$ and $ZrO_2$ films. The structure of $TiO_2$ is tetragonal and the structure of $ZrO_2$ is cubic. As a consequence, according to the above reasoning, $s <n>_{th}$ will be larger for $ZrO_2$ than for $TiO_2$ because its surface has a greater number of atoms [13].

Comparing the behaviour of different structures of the same compound, such as polycrystalline and amorphous $ZrO_2$, since the roughness of the surface is greater for the amorphous structure, the value of $Sc$ will be smaller and $M_J$ will be greater in the low fluence regime, even though the threshold fluence is the same for both types of crystals [13].

It is possible to extend this finding to the interpretation of the results of experiments designed to study the effect of the grain size on the ablation process, using a thin layer of Cd deposited on the surface of grains of sand [14]. In that work the area of the ablated surface is the same for the samples with different grain sizes, but those with grains larger than 1 mm have a lower quantity of particles than the samples with a grain size of 0.38 mm. Also, the sample with the larger grain size have much more defects on the surface than the sample with the smaller grains and, according to Eq. 4, will result is a greater value of $M_J$ although it will probably have a greater *rsd*.

### b. $M_J$ as a function of $E_d$ and $E_v$.

In this section we analyse the behaviour of Eq. 4 with $E_d$ and $E_v$, assuming that it could be different for metals than for crystals.

The behaviour of $M_J(f)$ is determined by the dependence of $A_1$ and $A_2$ on $E_v$ and $E_d$, respectively. Thus, according to Eq. 5, $A_1$ increases with fluence up to the asymptotic value $A_r$ and therefore, the maximum value of ejected material in the vaporization region is independent of the value of $E_v$. However, $E_v$ affects both the rate of change of $M_J$ with $f$ and its absolute value so that a larger $E_v$ results in a higher slope and a smaller yield. Numerical calculations exemplifying this behaviour with, $S_c = 30$ and $E_d = 30$ J.cm$^3$ are shown in Fig. 4.

As mentioned before, Eq. 5 is valid only for metals and organic solids, but not for ionic crystals, that are controlled by the lattice energy. As a result, $A_1$ loose physical sense and has to be changed for $A_2$, which makes the process in the low fluence regime also controlled by $E_d$ and $L_p$ (Eq. 6).

According to Eq. 6, $A_2$ increases monotonically with fluence from the initial value $n_m L_p$. Representative calculations are presented in Fig. 5, with constant values of $E_v = 30$ kJ.mol$^{-1}$, $S_c = 30$ and $L_p = 50$ nm. Since $E_d$ reflects the cohesion energy of the crystal structure, the amount of material ejected, at the same fluence, is larger for the solid with the smaller value of $E_d$, as expected. Then, $E_d$ sets a limit to the depth of penetration of the laser.

Based on these considerations, the magnitude of the difference between both regimes, vaporisation and ablation, should be determined by the geometrical characteristics of the crystal and the value of $E_d$. The solid with lower $E_d$ will produce the maximum amount of ablated material. A clear example of this behaviour is provided by the ablation of alkaline metal atoms [12]. Metallic Ba presents a body centred structure which has 68 % of occupied space while a surface with cubic compacted package (such as Ca and Sr) has 74% of space occupied. This implies a much higher value of $E_d$ for Ca and Sr than Ba and consequently, Ba should yield more ejection of material, in agreement with the reported results [12].

In the case of the laser threshold for film damage the upper limit to $M_J$ will be determined by the film depth $L_P$ regardless of the type of structure. The value of $E_v$ will determine the lower limit and the transition between both regimes will be governed by the structure of the crystal, in agreement with the experimental results [5].

The values of $A_1$ and $A_2$, that depend on the energy terms $E_v$ and $E_d$ respectively, determine the minimum (vaporization) and maximum (ablation) amount of ablated material that can be obtained, but at a given fluence, $M_J$ depends on $Sc$ as shown in Fig. 3.

**Conclusions**

The ejection of material from a solid target by the action of a laser pulse as a function of incident fluence shows three different regimes, according to the experimental and theoretical evidences available: vaporization, at low fluences, transition, at intermediate values and true ablation, at energies higher than the threshold, $f_{th}$. The whole range is well described by an empirical expression that provides a link between the low and high fluence limits. These limits are modeled by expressions reported in the literature: the low one by a typical Arrhenius equation dependent on $E_v$ and the high one, that properly is the ablation process, by a Beer´s law dependent on $E_d$ [3]. The intermediate region depends on a structural constant, *Sc*, and is consequently mainly governed by the structure of the crystal that determines the threshold to enter into the ablation regime. The value of *Sc* is representative of the various types of materials, so that *Sc* = 1-10 corresponds to metals or ionic crystals with a polished surface and *Sc* $\geq$ 20 are typical of ionic cleaved surfaces. In addition, the value of *Sc* has a remarkable effect on the shape of the curve of $M_J$ as a function of fluence: as Sc increases, the slope of the transition region increases and the smooth change observed at low values changes to an abrupt transition from the vaporization to the ablation regimes.

Even though the model is strictly applicable to single pulse results, it can also be used to qualitatively account for multipulse experiments.

Therefore, the model provides new elements, especially on the structure of the surface, that could be used in the numerical modeling which, to the best of our knowledge, have not been taken into account so far.

The simplicity of Eq. 4 could provide a powerful tool for the fast characterization of the various surfaces and the design of films according to their structure and the requirements of laser energy damage.

The model can also be extended to provide an explanation of the effect of the laser wavelength on the ablation process that will be the subject of a forthcoming communication.

Acknowledgments: The authors knowledge CONICET, ANPCYT, ACC and Fundación Antorchas for financial support.

**Figure Captions**

Fig. 1: Scheme of a topographic surface of a hypothetical crystal before and after the laser pulse.

Fig. 2: $M_J$ (Eq. 5) as a function of the fluence, $\phi$, with three different values of the surface coefficient: ($\triangle$) $S_c = 10$, (O) $S_c = 50$, (■) $S_c = 100$) and with $\phi_{th} = 30$ J.cm$^{-2}$, $E_d = 30$ J.cm$^{-3}$, $E_v = 38$ kJ.m$^{-1}$ and $T_o = 300$ K.

Fig. 3: $M_J$ (Eq. 5) as a function of the surface coefficient $S_c$ at four different fluences: (——) $\phi = \phi_{th}$, ($-\cdot\cdot-$) $\phi = 0.25 \times \phi_{th}$, (....) $\phi = 0.9 \times \phi_{th}$, (- - -) $\phi = 4 \times \phi_{th}$ and with $\phi_{th} = 30$ J.cm$^{-2}$, $E_d = 30$ J.cm$^{-3}$, $E_v = 38$ kJ.m$^{-1}$ and $T_o = 300$ K.

Fig. 4: $M_J$ (Eq. 5) as a function of the fluence, $\phi$, with $E_d = 30$ J.cm$^{-3}$ constant and three different values of $E_v$: (O) $E_v = 10$ J.cm$^{-2}$, ($\triangle$) $E_v = 20$ J.cm$^{-2}$ and ($\square$) $E_v = 30$ J.cm$^{-2}$

Fig. 5: $M_J$ (Eq. 5) as a function of the fluence, $\phi$, with $E_v = 30$ J.cm$^{-2}$ constant and four different values of $E_d$: ($\triangle$)$E_d = 0.01$ J.cm$^{-3}$, (O) $E_d = 0.1$ J.cm$^{-3}$, ($\square$) $E_d = 10$ J.cm$^{-3}$ and ($\blacklozenge$)$E_d = 30$ J.cm$^{-3}$.

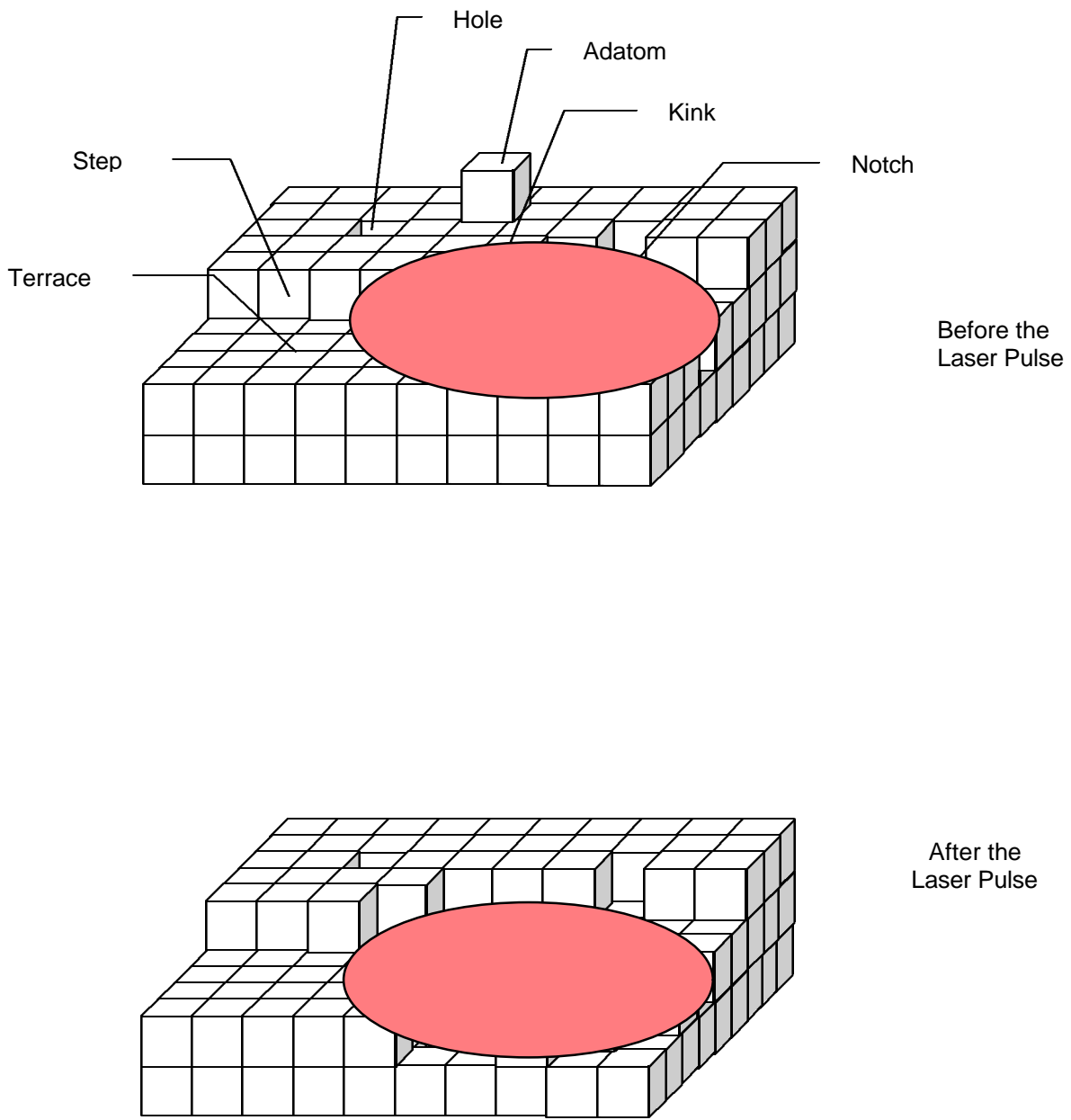

Figure 1

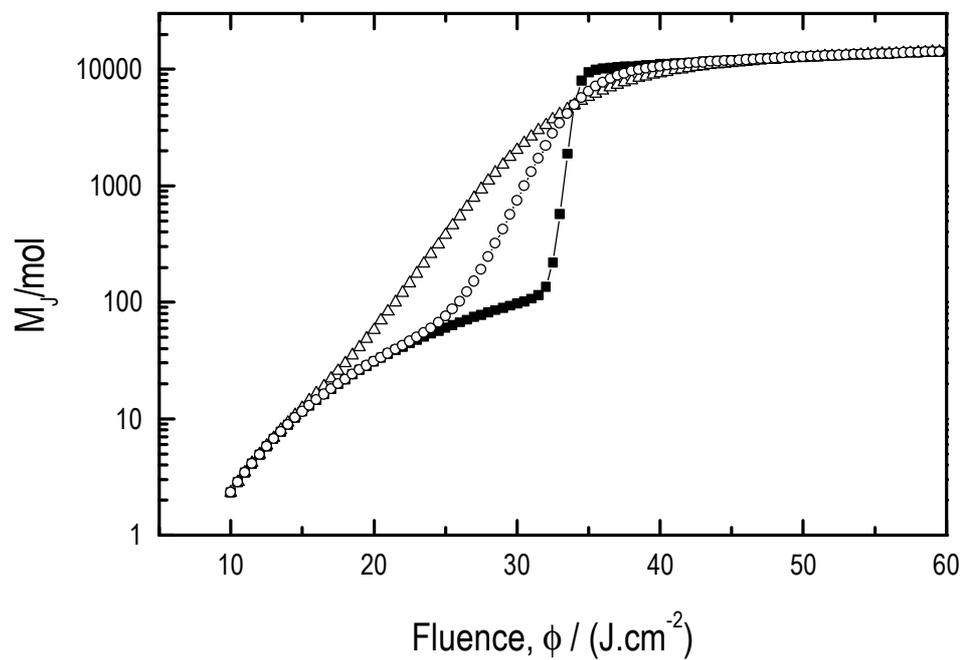

Figure 2

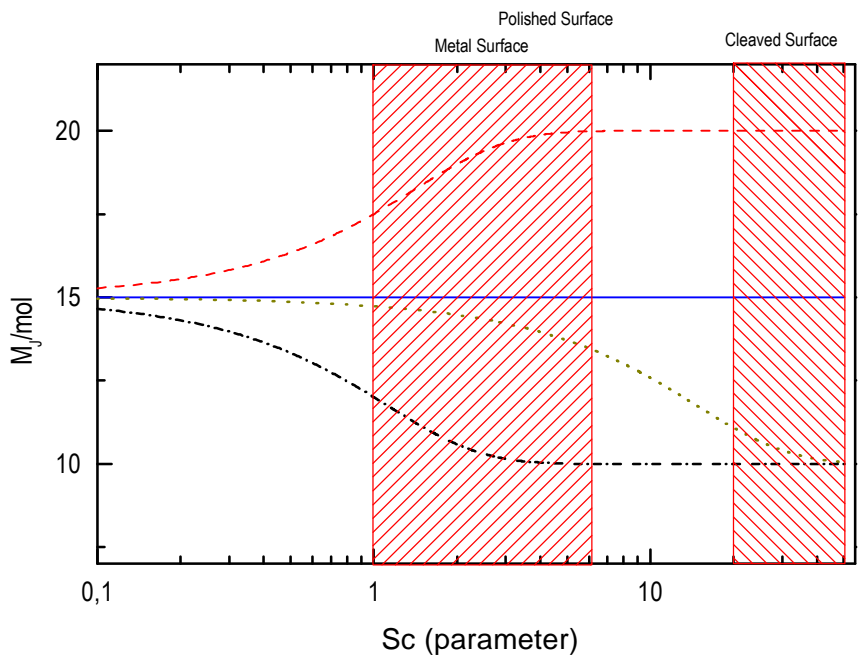

Figure 3

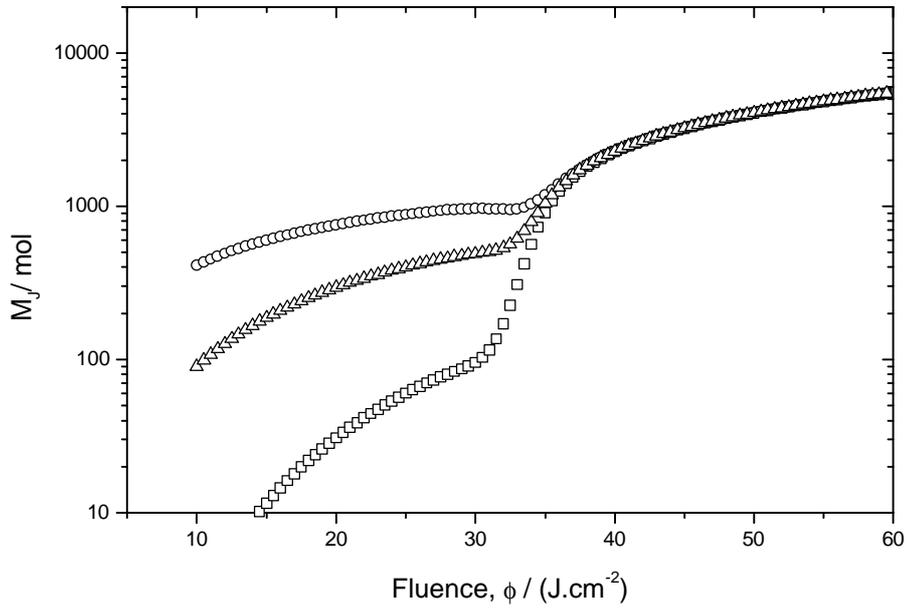

Figure 4

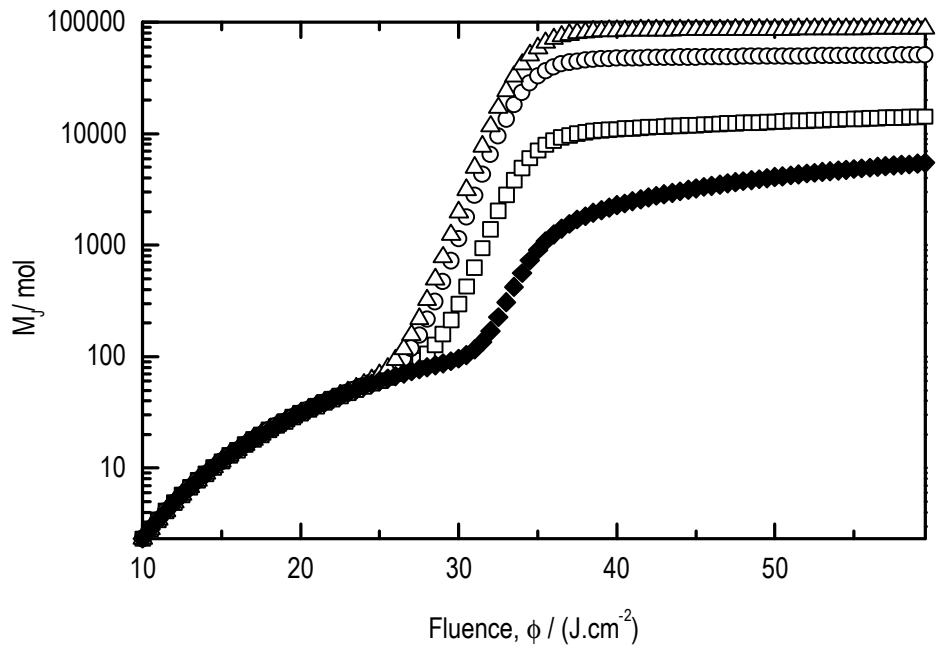

Figure 5